\documentstyle[aps,12pt]{revtex}  

\begin{document}
\def\nue{{\nu_e}}
\def\nuebar{{\bar{\nu}_e}}
\def\numu{{\nu_\mu}}
\def\numubar{{\bar{\nu}_\mu}}
\def\nutau{{\nu_\tau}}
\def\nutaubar{{\bar{\nu}_\tau}}
\def\nualpha{{\nu_\alpha}}
\def\nualphabar{{\bar{\nu}_\alpha}}
\def\nubeta{{\nu_\beta}}
\def\nubetabar{{\bar{\nu}_\beta}}
\def\nus{{\nu_s}}
\def\nusbar{{\bar{\nu}_s}}
\def\MeV{{\rm MeV}}
\def\la{\mathrel{\mathpalette\fun <}}
\def\ga{\mathrel{\mathpalette\fun >}}
\def\fun#1#2{\lower3.6pt\vbox{\baselineskip0pt\lineskip.9pt
  \ialign{$\mathsurround=0pt#1\hfil##\hfil$\crcr#2\crcr\sim\crcr}}}
\title{\ \ \ \\\ \\\ \\
Chaotic Amplification of Neutrino Chemical Potentials
by Neutrino Oscillations in Big Bang Nucleosynthesis\\}
\vskip 0.4in
\author{Xiangdong Shi}

\address{Department of Physics, Queen's University, Kingston,
Ontario K7M 5W9, Canada}

\maketitle
\vskip 1in
\centerline{ABSTRACT}
\vskip 0.2in
\begin{abstract}                
We investigate in detail the parameter space of active-sterile neutrino
oscillations that amplifies neutrino chemical potentials
at the epoch of Big Bang Nucleosynthesis. We calculate the
magnitude of the amplification and show evidences of chaos in the
amplification process. We also discuss the implications
of the neutrino chemical potential amplification in the Big Bang
Nucleosynthesis. It is shown that with
a $\sim 1$ eV $\nue$, the amplification of its chemical potential
by active-sterile neutrino oscillations
can lower the effective number of neutrino species
at Big Bang Nucleosynthesis to significantly below 3.
\end{abstract}
\newpage
\section{INTRODUCTION}
Neutrino oscillations have been suggested to explain several experimental
results and if proven true, they will represent a significant step toward
physics beyond the standard model of particle physics \cite{Langacker}.
Mixings between active neutrinos ($\nue$, $\numu$ or $\nutau$) and
sterile neutrinos (hypothetical neutrinos that do not interact with
known particles via the strong, weak or electromagnetic interactions) are
one possible source of neutrino oscillations.
The most stringent constraints on the parameters of
active-sterile neutrino mixings come from
cosmological and astrophysical considerations
\cite{Dolgov,Enqvist,Cline,Shi,Sigl}.
In particular, the active-sterile neutrino oscillations at the epoch
of Big Bang Nucleosynthesis (BBN) have been investigated extensively and
tight constraints have been obtained based on the primordial
$^4$He abundance in our universe. Interestingly, it was recently pointed out
by Foot, Thomson and Volkas \cite{FTV}
that some parameter space of the active-sterile neutrino mixings can
amplify neutrino asymmetries (neutrino chemical potentials)
so that the previous constraints on active-sterile
neutrino mixings based on BBN can be alleviated \cite{FTV,Foot}.
Also an asymmetry in the electron
neutrino sector as large as $\sim 0.1$ at the time of BBN can
change significantly the BBN prediction of the primordial $^4$He abundance.

In section 2 of this paper we expand the original investigation of Foot,
Thomson and Volkas \cite{FTV},
by calculating the parameter space that amplifies neutrino
chemical potentials and the magnitude of the amplification. But instead of
relying on a simplified equation that only applies outside the resonant
regime and when neutrinos are incoherent, we analyze the problem based
on the original equations in the density matrix formalism, both analytically
and numerically. Our analyses reveal many interesting features of the
amplification process that cannot be revealed by the simplified approach.
For example, the neutrino asymmetry can be oscillatory long after the
initial resonant crossing. There are evidences which suggest that
the oscillatory asymmetry is chaotic. As a result,
although the order of magnitude of the final neutrino chemical potential
is readily predictable, the sign of the chemical potential is very sensitive
to the mixing parameters and the input parameters of numerical calculations.
This oscillatory behavior and evidences of a chaotic amplification
are probed in section 2.

In section 3, we discuss two implications of our results in section 2 and
show how active neutrinos as dark matter candidates--having $\sim 1$eV mass--
can lower the effective number of neutrino species in BBN to significantly
below 3 by mixing with a lighter sterile neutrino.  

\section{Formalism and Calculations}

Through out the paper, we adopt a unit in which
$\hbar=c=k=1$. We also use a convention to denote
the number density of a particle $i$ by $N_i$, and the number density
relative to its equilibrium value ($2\zeta (3)T^3/\pi ^2$ for photons,
3/4 of that for electron-positrons, and 3/8 of that for neutrinos) by $n_i$.
The Hubble expansion rate is $H=7T^2/M_p$ where $M_p$ is the Planck mass
and $T$ the temperature of the universe.
Finally, $T_6$ denotes $T$ in the unit of MeV.
Since we are only concerned with the era of BBN, we limit
our discussion to $1\la T_6\la 100$.

Mixtures of an active neutrino $\nualpha$ and a sterile neutrino $\nus$ can
be described by a density matrix \cite{Enqvist}
\begin{equation}
\rho_\nu=\left(\begin{array}{cc}
\rho_{\alpha\alpha} & \rho_{\alpha s}\\
\rho_{s\alpha} & \rho_{ss}\end{array} \right)
={P_0I+{\bf P}\cdot {\bf \vec{\sigma}}\over 2},
\label{matrix}
\end{equation}
where {$\vec{\sigma}$} are the Pauli matrices.
The number densities of $\nualpha$ and $\nus$ in the mixture,
relative to their equilibrium values, are respectively
\begin{equation}
n_{\nu_\alpha}={P_0+P_z\over 2},\quad\quad n_{\nu_s}={P_0-P_z\over 2}.
\end{equation}
The evolution of the total relative number density of the neutrino
mixture at the epoch of BBN is \cite{Enqvist}
\begin{equation}
\dot P_0=\sum_{i=e,\nu_\beta; \beta\neq\alpha}
\bigl\langle\Gamma(\nualpha\nualphabar\rightarrow i\bar i)\bigr\rangle
(n_i n_{\bar i}-n_\nualpha n_{\nualphabar}),
\label{totnum}
\end{equation}
where $\langle\Gamma\rangle$ are reaction rates averaged
over a thermal spectrum.
Values of $\langle\Gamma\rangle$ are listed in table 1 of ref.\cite{Enqvist}
or ref. \cite{Shi}. The evolution of {\bf P} is \cite{Enqvist}
\begin{equation}
\dot {\bf P}={\bf V}\times{\bf P}+\dot P_0{\bf\hat z}-D{\bf P_\bot}
\label{master}
\end{equation}
where {\bf V} represents the frequency and the axis of the oscillation in
the {\bf P}-space, and {\bf P}$_\bot=P_x{\bf\hat x}+P_y{\bf\hat y}$.
The $D$-term represents the damping of {\bf P}$_\bot$
due to neutrino interactions which constantly reduce a mixed neutrino
state into an eigenstate of either $\nualpha$ or $\nus$.

At the epoch of BBN,
\begin{equation}
V_x={\delta m^2\over 2E}\sin 2\theta,\quad
V_y=0,\quad
V_z=-{\delta m^2\over 2E}\cos 2\theta+V_\alpha^L+V_\alpha^T,
\end{equation}
where $\delta m^2$ and $\theta$ are the usual vacuum mixing parameters, and
$E$ is the energy of the neutrinos. $V_\alpha^L$ is the contribution of
the matter effect from asymmetries in the background plasma \cite{Raffelt}:
$$V_\alpha^L=\sqrt{2}G_FN_\gamma\Bigl\{L_0
                  +0.375\Big[2(n_{\nu_\alpha} -n_{\nualphabar})+\sum_
                  {\nubeta\neq\nualpha}(n_\nubeta-n_\nubetabar)\Bigr]\Bigr\}$$
\begin{equation}
	  \approx 0.13 G_FT^3\Bigl[8L_0/3
                  +2(n_{\nu_\alpha} -n_{\nualphabar})+\sum_
                  {\nubeta\neq\nualpha}(n_\nubeta-n_\nubetabar)\Bigr]
\label{VL}
\end{equation}
where $L_0$ represents the contributions from the baryonic asymmetry
as well as the asymmetry in electron-positions, and is $\sim 10^{-9}$.
$N_\gamma$ is the photon number density. The $n_\nu-n_{\bar\nu}$ terms
represent the asymmetries in active neutrinos and thus their non-zero
chemical potentials. If $\xi_\nu$--the chemical potential of $\nu$ divided by
$kT$--is much smaller than 1, $n_\nu-n_{\bar\nu}\approx 1.8\xi_\nu$.

$V_\alpha^T$ is the contribution of the matter effect due to
a finite temperature \cite{Raffelt}:
\begin{equation}
\begin{array}{rll}
V_\alpha^T=&-\sqrt{2}G_FN_\gamma\Bigl[12.61ET(n_\nualpha+n_\nualphabar)/4M_Z^2
                       +12.61ET/M_W^2\Bigr],&\quad\alpha =e;\cr
     =&-\sqrt{2}G_FN_\gamma\Bigl[
       12.61ET(n_\nualpha+n_\nualphabar)/4M_Z^2\Bigr],&\quad\alpha=\mu,\tau.
\end{array}
\label{VT}
\end{equation}
It has been shown that eq.~(\ref{totnum})--(\ref{VT}) give a good
description of neutrino oscillations in BBN if the average neutrino energy
$E\approx 3.151T$ is inserted in {\bf V} and if $D$ is thermally averaged
\cite{Enqvist}. Therefore, numerically,
\begin{equation}
\begin{array}{rll}
V_\alpha^T\approx &-250G_F^2T^5,&\quad\quad\alpha =e;\cr
          \approx &-70G_F^2T^5, &\quad\quad\alpha=\mu,\tau.
\end{array}
\label{VTnum}
\end{equation}
The damping coefficient $D$, consisting of contributions from both
elastic scatterings and inelastic scatterings of $\nualpha$,
is \cite{Enqvist,Shi}
\begin{eqnarray}
D&\approx (1.3+0.4n_\nualpha+0.5n_\nualphabar)G_F^2T^5,
  \quad \alpha=e;\\\nonumber
 &\approx (0.8+0.4n_\nualpha+0.5n_\nualphabar)G_F^2T^5,
  \quad\alpha=\mu,\tau.
\end{eqnarray}
The reason we leave out $n_\nualpha$ and  $n_\nualphabar$ explicitly
without approximating them to 1 is for the convenience of calculating
the difference in the coefficient between the $\nualpha$-$\nus$ and
$\nualphabar$-$\nusbar$ oscillations. Since we often compare $D$ to
the Hubble expansion rate $H$, we note
\begin{equation}
D\approx 0.5T_6^3H.
\end{equation}

The initial condition for eq.~(\ref{totnum}) and (\ref{master})
is usually chosen to be $P_0=P_z=1$ and $P_x=P_y=0$, at $T_{\rm init}$
when $V_\alpha^T$ dominates over $\delta m^2/2E$, i.e.,
\begin{equation}
T_{\rm init}\gg 
15\left\vert{\delta m^2\cos 2\theta\over {\rm eV}^2}\right\vert^{1/6}
\label{Tinit}
\end{equation}
(for the moment we assume any neutrino asymmetry is negligible).
That is, the neutrino ensemble consists purely
of $\nualpha$, which is a good approximation because $V_z\gg V_x$ so that
{\bf V} is almost aligned with the ${\bf\hat z}$-axis.

As the universe expands and its temperature drops,
$\vert V_x/V_z\vert$ becomes larger,
the amplitude of the oscillation consequently increases. Eventually,
if there is no amplification of neutrino asymmetries, $V_\alpha$
becomes negligible, and {\bf V} settles down into its vacuum value.
During the process, if the mixing has $\delta m^2<0$
($\nualpha$ heavier than $\nus$), a resonance
can occur when {\bf V} crosses the ${\bf\hat x}$-axis at a temperature
\begin{equation}
T_{\rm res}\approx 13 (16)\cdot
\left\vert{\delta m^2\cos 2\theta\over 1{\rm eV}^2}\right\vert^{1/6}{\rm MeV}
\quad\quad {\rm for\ } \alpha=e{\rm \ } (\mu,\tau).
\end{equation}
{\bf P} and {\bf V} before and after the resonance are illustrated in
figure 1.

During the oscillation, the interactions between $\nualpha$ and the background
plasma play two roles. First, the interactions (including both elastic and
inelastic ones, represented by the $D$-term) reduce mixed neutrino states
into either $\nualpha$ or $\nus$, which effectively damp the amplitude of
${\bf P}_\bot$ and at the same time randomize the phase of the neutrino
oscillation. When the regenerated $\nualpha$ (and $\nus$ but mostly $\nualpha$)
oscillate into $\nus$ (and $\nualpha$) again, the portion of
$\nualpha$ in excess of $\nus$, $P_z$, decreases toward 0.
Secondly, the inelastic process--$\nualpha\nualphabar$ pair productions
(the $\dot P_0{\bf\hat z}$ term)--constantly replenishes the number of
$\nualpha$ that is being depleted by oscillation, maintaining its population
as a full relativistic species as long as such pair productions are potent
($T\ga 3 $ MeV for $\nue$ and $\ga 5$ MeV for $\numu$ or $\nutau$).

Eq.~(\ref{matrix})--(\ref{Tinit}) can be equally applied to the
anti-neutrino sector, with notations for particles and anti-particles
switched and $L_0$ replaced by $-L_0$. Apparently, since $\nualpha$ and
$\nualphabar$ can only be produced as pairs, $\dot P_0={\dot{\bar P_0}}$.

The relevance to BBN comes at $T\sim 1$ MeV when the neutron to proton ratio
freezes out. If a significant population of $\nus$ is produced, or a
significant asymmetry in $\nue\nuebar$ is generated through the active-sterile
neutrino oscillations, the neutron to proton ratio can be affected and the
resultant $^4$He primordial abundance altered from the standard BBN
predictions. When neutrino asymmetries are negligible, to be consistent with
the observed primordial $^4$He abundance requires \cite{Enqvist,Shi}
\begin{equation}
\delta m^2 \sin^42\theta\la 10^{-9}{\rm eV}^2
\label{constraint1}
\end{equation}
(The bound on the $\nue$-$\nus$ mixing
is tighter on the low $\delta m^2$ end. See
refs. \cite{Enqvist} and \cite{Shi} for precise constraints.)

Under conditions that
\begin{equation}
\vert{\bf V}\vert\gg D\gg\vert\dot{\bf V}\vert/\vert{\bf V}\vert,
\label{condition}
\end{equation}
i.e., the damping of {\bf P} and the change in {\bf V} are negligible
within one cycle of oscillation of {\bf P},
eq.~(\ref{master}) can be simplified  to the lowest order to
\begin{equation}
P_x=V_xP_z/V_z,\quad P_y=0,\quad\dot P_z=-DV_x^2P_z/(V_x^2+V_z^2)+\dot P_0.
\label{simple}
\end{equation}
In the epoch of our concern, and in the absence of an amplification of
neutrino asymmetries,
eq.~(\ref{condition}) is satisfied except near the resonance region where
$\vert V_z\vert\sim\vert V_x\vert$. This is because
$V_\alpha^T={\cal{O}}(10^2)D$ and
$\vert\dot V_\alpha /V_\alpha\vert\sim H\ll D$ at $T_6\ge 1$. We
will discuss the case of amplified asymmetries later in the section.

Similarly for anti-neutrinos, the approximate equations are
\begin{equation}
\bar P_x=V_x\bar P_z/\bar V_z,\quad \bar P_y=0,\quad
{\dot{\bar P_z}}=-\bar DV_x^2\bar P_z/(V_x^2+\bar V_z^2)+{\dot{\bar P_0}}
\label{simplebar}
\end{equation}
under conditions that $\vert{\bf\bar V}\vert\gg\bar D
\gg\vert\dot{\bf\bar V}\vert/\vert{\bf\bar V}\vert$.
Assuming there is no asymmetries in neutrinos other
than the oscillating $\nualpha\nualphabar$ sector, 
\begin{equation}
\bar V_z=V_z-2\beta(\Delta P_z+8L_0/3)=V_0-\beta(\Delta P_z+8L_0/3)
\label{VVbar}
\end{equation}
where $\beta=0.375\sqrt{2}G_FN_\gamma\approx 0.13G_FT^3$,
$V_0=-\delta m^2\cos 2\theta/2E+V_\alpha^T$ and $\Delta P_z=P_z-\bar{P_z}
=2(n_\nualpha-n_\nualphabar )$. It is also noted that
$\bar D-D\approx 0.05\Delta P_zG_F^2T^5\ll D\Delta P_z$.

The asymmetry in the $\nualpha\nualphabar$ sector can then be described
by $\Delta P_z$ which to its lowest order satisfies
$$\dot{\Delta P_z}=
  DP_zV_x^2\Bigl({1\over V_x^2+\bar V_z^2}-{1\over V_x^2+V_z^2}\Bigr)
 -D{V_x^2\over V_x^2+\bar V_z^2}\Delta P_z
 +(\bar D-D){V_x^2\over V_x^2+ \bar V_z^2}\bar P_z$$
\begin{equation}
      \approx{DV_x^2\over V_x^2+[V_{0}-\beta(\Delta P_z+8L_0/3)]^2}
        \left\{{4V_{0}\beta (\Delta P_z+8L_0/3)P_z\over
  \{V_x^2+[V_{0}+\beta (\Delta P_z+8L_0/3)]^2\}\Delta P_z}-1\right\}\Delta P_z.
\label{DeltaPz}
\end{equation}
The equation resembles equation~(15) of Foot, Thomson and Volkas
\cite{FTV} except that their equation omitted the second term and $L_0$, and
assumed $P_z$ to be 1.

When $V_0<0$, eq.~(\ref{DeltaPz}) is a damping equation for
$\vert\Delta P_z\vert >8L_0/3$, and no amplification of $\vert\Delta P_z\vert$
to $\gg 10^{-9}$ can occur. This rules out $\nualpha$-$\nus$ mixings with
$\delta m^2>0$ ($\nus$ heavier than $\nualpha$). Only when
$\delta m^2<0$ and $V_0$ switchs to a positive value (resonance crossing)
before $T\sim 1$ MeV does
an amplification of $\Delta P_z$ become plausible.

For mixings with $\delta m^2<0$, $V_0$ still starts out negative at high
temperatures $\gg T_{\rm res}$. Any $\vert\Delta P_z\vert >8L_0/3$ will be
damped toward an asymptotic value such that $\dot{\Delta P_z}\rightarrow 0$.
Thus
\begin{equation}
\Delta P_z\rightarrow
    -{8\over 3}L_0+\left\vert{8V_x^2L_0/3\over 4V_0\beta P_z}\right\vert
\approx -{8\over 3}L_0
\label{residue}
\end{equation}
when $\vert V_0\vert\gg \vert V_x\vert$. 

When the system enters the resonant regime,
eq.~(\ref{DeltaPz}) does not apply. We have to go back
to the original equation~(\ref{master}) and its anti-neutrino counterpart,
which give
\begin{equation}
\begin{array}{rl}
\dot{\Delta P_x}=&
     -V_0\Delta P_y-\beta(\Delta P_z+8L_0/3)(P_y+\bar P_y)-D\Delta P_x,\cr
\dot{\Delta P_y}=&
      V_0\Delta P_x+\beta(\Delta P_z+8L_0/3)(P_x+\bar P_x)
      -V_x\Delta P_z-D\Delta P_y,\cr
\dot{\Delta P_z}=&V_x\Delta P_y.
\end{array}
\label{resonance}
\end{equation}
where $\Delta P_x=P_x-\bar P_x$ and  $\Delta P_y=P_y-\bar P_y$.
If we don't want $\nus$ to be brought into equilibrium, we have to restrict
our discussions to the parameter space that satisfies
eq.~(\ref{constraint1}). Then: (1) the
resonance crossing is non-adiabatic; (2)
$D$ has to be small enough compared to the time scale of the
resonance crossing so that most of the neutrinos do not scatter
during the crossing. The non-adiabatic change in {\bf V} leads to a coherent
oscillation of {\bf P} around the new {\bf V} (see figure 1(b)).
The impotency of scatterings enables this coherency to
be maintained throughout the resonant regime and beyond until 
$D\ga \vert\dot {\bf V}\vert/\vert{\bf V}\vert$ so that the interactions
have enough time to randomize the phases of neutrinos again. It also implies
that the $D$-term in eq.~(\ref{resonance}) can be dropped in resonance.

When the temperature was high above $T_{\rm res}$ so that eq.~
(\ref{DeltaPz}) applies, $\Delta P_x\approx V_x\Delta P_z/V_z$,
$\Delta P_y$ and $\dot{\Delta P_y}$ are approximately 0. But in and right after
resonance, the approximation breaks down because of the
rapid change of $V_z$ and $\bar V_z$. Instead $\Delta P_y$ becomes oscillatory
with a frequency of $\sim V_z$ (figure 1(b)).
$\vert\Delta P_z+8L_0/3\vert$ will quickly
be of order $10^{-9}$, but a more important question is whether
$\vert\Delta P_z+8L_0/3\vert$ (and thus $\vert\Delta P_z\vert$) can be
amplified to $\gg 10^{-9}$. Assuming that $\vert\beta\Delta P_z\vert\ll V_0$,
the amplitude of the oscillating $\Delta P_y$ will be of order
$\sim \vert P_zV_x(V_z^{-1}-\bar V_z^{-1})\vert\approx 
\vert 2V_x\beta\Delta P_z/V_0^2\vert$ (from now on $P_z\sim 1$ and
$L_0$ is dropped for simplicity because
$\vert\Delta P_z+8L_0/3\vert\sim\vert\Delta P_z\vert$).
The amplification of $\vert\Delta P_z\vert$ depends on
whether $\dot{\Delta P_z}$ has enough time to change $\Delta P_z$ by a
factor of more than 1, i.e.,
\begin{equation}
\left\vert{2V_x^2\beta\over V_0^2}\right\vert\ga V_0.
\label{amplifi}
\end{equation}
Since $V_0$ is a changing quantity, the condition of amplification depends
on which $V_0$ to choose. For a crude estimate, a reasonable
choice is the $V_0$ at the time when $\Delta P_y$ oscillates one cycle
since the resonance (so that eq~(\ref{amplifi}) is meaningful). So
\begin{equation}
V_0\sim \dot{V_0}\cdot V_0^{-1}\sim {H\over V_0}\cdot{\vert
\delta m^2\vert\over 2E}.
\end{equation}
Solving the equation assuming $E=3.151T$ yields
$V_0\sim 10^{-2}\vert\delta m^2\vert^{-0.25}\cdot\vert\delta m^2/2E\vert$. Thus
the condition of amplifying $\Delta P_z$, i.e., eq.~(\ref{amplifi}), is
\begin{equation}
\vert\delta m^2\vert^{-5/12}\sin^2 2\theta\ga 10^{-12}.
\label{boundary}
\end{equation}

The growth of $\vert\Delta P_z\vert$ is limited once
$\vert\beta\Delta P_z\vert\gg V_0$ (when $\vert\Delta P_z\vert \gg
\vert\delta m^2/9{\rm eV}^2\vert T_6^{-4}$). Because at this moment,
the amplitude of $\Delta P_y$ becomes $\sim \vert 2V_x/\beta\Delta P_z\vert$,
and the amplitude of $\dot{\Delta P_z}/\Delta P_z$ becomes
$\sim 2V_x^2/\beta\Delta P_z^2$, proportional to $\Delta P_z^{-2}$.

An interesting feature of $\Delta P_z$, shown both from eq.~(\ref{amplifi})
and numerical calculations, is that it keeps oscillating below the
resonant temperature (figure 2). This is a direct consequence of
the coherent oscillation of {\bf P} and $\bar{\bf P}$.
As a result of the oscillation, the sign of
$\Delta P_z$ flips (so does $V_z$ and $\bar {V_z}$)
unless the change of $\Delta P_z$ in each cycle
is smaller than the amplitude of $\Delta P_z$ itself, i.e.,
\begin{equation}
\vert\Delta P_z\vert >{V_x^2\over\vert\beta\Delta P_z\vert}\cdot
\vert\beta\Delta P_z\vert^{-1}.
\end{equation}
Since for parameters that satisfy eq.~(\ref{boundary}), the amplitude of
$\Delta P_z \ga (\vert\delta m^2\vert/9{\rm eV}^2)T_6^{-4}$,
a rough realization of the condition that $\Delta P_z$ will not
be oscillatory between positive and negative values is
\begin{equation}
{\vert\delta m^2\vert\over \sin^2 2\theta}\ga 9T_6^4{\rm eV}^2.
\label{stillosc}
\end{equation}
At $T_6=1$, $\Delta P_z$ will not be oscillatory if
$\vert\delta m^2\vert/\sin^2 2\theta\ga 9{\rm eV}^2$.

Once $\Delta P_z$ stops flipping its sign,
$\vert\dot{\bf V}\vert/\vert{\bf V}\vert$ and
$\vert\dot{\bar{\bf V}}\vert/\vert\bar{\bf V}\vert$
decrease dramatically so that the damping $D$-term becomes more and more
important. Eventually eq.~(\ref{DeltaPz}) reapplies. Assuming $P_z\sim 1$ 
and neglecting $L_0$ and $V_x$ in the denominator eq.~(\ref{DeltaPz}) yields
\begin{equation}
\dot{\Delta P_z}\approx D{V_x^2\over (V_0-\beta\Delta P_z)^2}
\left[{4V_0\beta\over (V_0+\beta\Delta P_z)^2}-1\right]\Delta P_z.
\end{equation}
Since $V_0>0$ after the initial resonance,
this is an amplification equation if
\begin{equation}
{4V_0\beta\over (V_0+\beta\Delta P_z)^2}>1,\quad{\rm or}\quad
\delta m^2\la 10^2{\rm eV}^2.
\label{constraint2}
\end{equation}
The amplification will not stop until
\begin{equation}
D{V_x^2\over (V_0+\beta\Delta P_z)^2}
{4V_0\beta\over (V_0-\beta\Delta P_z)^2}\la H,
\end{equation}
which occurs at $\vert\beta\Delta P_z\vert\ga V_0\sim \vert\delta m^2/2E\vert$
and
\begin{equation}
\left({18T_6^7\sin^22\theta\over\vert\delta m^2\vert}\right)
\left({\delta m^2/2E\over \beta\Delta P_z}\right) ^4\la 1.
\end{equation}
At $T_6\sim 1$, nearly all the mixing parameters that show
no oscillatory $\Delta P_z$ have $18T_6^7\sin^22\theta/\vert\delta m^2\vert
<1$, so $\vert\beta\Delta P_z\vert$ is limited to $\sim
\vert\delta m^2/2E\vert$ and
\begin{equation}
\vert\Delta P_z\vert\sim {\vert\delta m^2\vert\over 9T_6^4}
\label{DeltaPzest}
\end{equation}
at a temperature of $\sim 1$ MeV. This limit is confirmed
by our numerical calculations and is similar to that
of Foot, Thomson and Volkas\cite{FTV} based on their simplified equation.
For $\vert\delta m^2\vert\ga 10{\rm eV}^2$, however, since $\Delta P_z$ has
to be much smaller than $P_z\sim 1$, $\Delta P_z\sim 0.1$.

The oscillatory behavior of $\Delta P_z$ after resonance is illustrated
in figure 2 (a), (b) and (c), for three different $\nue$-$\nus$
mixing parameters. Each graph is the result of several millions steps of
integrations of eq.~(\ref{totnum}) and eq.~(\ref{master}) by
adaptive Runge-Kutta method, with
an error of less than 10$^{-10}$ in each step. In figure 2 (a), the mixing
parameters do not satisfy equation~(\ref{stillosc}) at $T_6=1$, so
$\Delta P_z$ is still oscillatory at 1 MeV.
In figure 2 (b) and (c), $\vert\Delta P_z\vert$ settles down to
$\sim\vert\delta m^2/9{\rm eV}^2\vert T_6^{-4}$ at $T_6\ga 1$, in line
with our estimate eq.~(\ref{DeltaPzest}).
Our numerical calculations also show that although the final
settle-down value of $\vert\Delta P_z\vert$ is predictable, the sign of
$\Delta P_z$ seems random among different parameter choices.  
For example, in case (b),
$\delta m^2=-10^{-2}$ eV$^2$ and $\sin^22\theta=10^{-4.25}$,
a small change of $\sin^22\theta$ to $10^{-4.1875}$
yields an opposite sign of the final $\Delta P_z$.
The sign of $\Delta P_z$ can also be flipped by slight changes
in the initial $L_0$ (as tiny as 0.01\%) and calculational parameters,
such as a different error control (from 10$^{-10}$
to 5$\times 10^{-10}$ in our example), or a step size, or even a slightly
different relation
between the average neutrino energy and the temperature (from $E=3.151T$ to
$E=3.150T$ in our example). This is due to the large number
of oscillations of $\Delta P_z$ before it approaches one of the two
possible values, so that the final $\Delta P_z$ is very sensitive to the
input and calculational parameters.
Such behavior is not so significant in the case of 
$\delta m^2=-1$ eV$^2$ and $\sin^22\theta=10^{-8}$, because the
number of oscillations of $\Delta P_z$ is small (figure 2 (c)). In this case,
using $E=3.150T$ the evolution of {\bf P} and $\bar{\bf P}$
traces the evolution of {\bf P} and $\bar{\bf P}$ using
$E=3.151T$ very well (in a sense that their difference is obviously
still a perturbation at lower temperatures). This is also true if
we change the initial $L_0$ by 0.01\% (although the resultant
perturbation can be several
percent in the oscillatory epoch of $\Delta P_z$).
Nevertheless, a small change of $\sin^22\theta$ to 10$^{-8.0625}$ still flips
the sign of the final $\Delta P_z$.

Figure 2 (a) and (b) suggest a chaotic behavior in the epoch
of oscillating $\Delta P_z$. Figure 2 (c) might be intrinsically chaotic too
but the time scale of the oscillatory epoch
may be too short for such behavior to show up.
To find more evidences of chaos, we try to determine the
Lyapunov exponents of the system\cite{chaos}.
Assuming $\vec\phi(t)=(\Delta P_x,\Delta P_y,\Delta P_z)$
represents a solution of eq~(\ref{resonance}), we investigate the behavior of
a nearby solution $\vec\phi(t)+\vec{\delta\phi} (t)=
(\Delta P_x,\Delta P_y,\Delta P_z)+(\delta P_x,\delta P_y,\delta P_z)$
($\delta P_x$ and $\delta P_y$ are assumed to arise from
$P_x$ and $P_y$ only so that we can carry out our analysis).
The evolution of the $\vec{\delta\phi}(t)$ is obtained by linearizing
eq.~(\ref{resonance}):
\begin{equation}
\dot{\vec{\delta\phi}}=M\vec{\delta\phi}
\end{equation}
where matrix
\begin{equation}
M=\left(\begin{array}{ccc}
-D                  &-V_0-\beta\Delta P_z&-\beta(P_y+\bar P_y)    \cr
-V_0-\beta\Delta P_z&-D                  &-V_x+\beta(P_y+\bar P_y)\cr
0                   &V_x                 &0\end{array}\right).
\end{equation}
The eigenvalue of $M$, called the Lyapunov exponent, is
\begin{equation}
\lambda={\beta(P_y+\bar P_y)(V_0+\beta\Delta P_z)\over 
         \beta(P_x+\bar P_x)-V_x}-D.
\label{eigenvalue}
\end{equation}
If $\lambda$ is positive, nearby solutions will depart exponentially
in phase space within a timescale of $\lambda^{-1}$. In our problem,
since $P_x+\bar P_x$, $P_y+\bar P_y$ and $\Delta P_z$ are all oscillatory,
so is $\lambda$. A crude analysis is to plug in the amplitudes of
$P_x+\bar P_x$, $P_y+\bar P_y$ and $\Delta P_z$, to see whether
$\lambda$ can be sometimes positive.
In the limiting case of $\vert\beta\Delta P_z\vert\ll V_0$,
$P_x+\bar P_x\sim P_y+\bar P_y\sim V_x/V_0$, the resultant $\lambda$
is $\sim\beta V_0/(\beta -V_0) -D$ which can certainly be positive
if $V_0 >0$ and $\vert\delta m^2\vert\ll 9T_6^4{\rm eV}^2$,
conditions satisfied after the initial resonance.
In the limiting case of $\vert\beta\Delta P_z\vert\gg V_0$,
$P_x+\bar P_x\sim P_y+\bar P_y\sim V_x/\beta\Delta P_z$, the resultant
$\lambda$ is $\sim \beta\Delta P_z/(1-\Delta P_z)-D$,
which can again be positive (remember $\Delta P_z\ll 1$ and $\beta\Delta P_z
\gg V_0\gg D$). Therefore, our crude analysis shows that $\lambda$ can
at least be positive within a timescale of order $\lambda^{-1}$
intermittently.\footnote{In the analysis,
we have chosen a particular set of nearby solutions, namely those having
deviations in $P_x$ and $P_y$ but not
in $\bar P_x$ and $\bar P_y$. The other extreme choices, in which $\delta P_x$
or $\delta P_y$ arises only in one particle population but not in its
anti-particles, merely change the sign of $\beta\Delta P_z$ in the nominator of
eq.~(\ref{eigenvalue}), thus do not affect the general behavior of $\lambda$.}
In other words, the system is not a classical
textbook example of a chaotic system.

We are ultimately interested in the region of mixing parameters that
amplifies neutrino chemical potentials, and the size of the amplification.
In figure 3 (a) and (b), we plot the
parameter space allowed by BBN that amplifies neutrino chemical potentials.
The boundary to the right
which excludes parameters that bring $\nus$ into equilibrium
is adopted from Shi, Schramm and Fields\cite{Shi}.
The boundary to the left that distinguishes parameters that amplify
neutrino chemical potentials from those that do not
is based on our numerical calculations (smoothed), and
agrees with our analytical estimate eq.~(\ref{boundary})
within an order of magnitude. The lower cut
on $\vert\delta m^2\vert$ is determined by requiring $T_{\rm res}\ge 1$ MeV.
The upper cut on $\vert\delta m^2\vert$ is dictated by laboratory bounds
on $\nue$ mass in the $\nue$-$\nus$ mixing case, and by 
eq.~(\ref{constraint2}) as well as cosmological
considerations \cite{MDM} in the $\numu$($\nutau$)-$\nus$ mixing case.
The boundary that singles out parameters that have oscillatory $\Delta P_z$ at
1 MeV is plotted according to eq.~(\ref{stillosc}) which is confirmed by
our numerical calculations.

We note that our numerical calculation of $\delta m^2=-1$eV$^2$,
$\sin^22\theta=10^{-8}$
$\nu_e$-$\nus$ mixing yields an opposite sign of $\Delta P_z$
from that of Foot, Thomson and Volkas\cite{FTV}. But this may not be
surprising due to the chaotic feature of the system, that different signs
of $\Delta P_z$ may arise from different initial $L_0$,
or even different choices of integrators,
different errors or step sizes. It is also noted that the simplified
equation in ref. \cite{FTV} (corresponding to eq.~(\ref{DeltaPz}) without
the second term and $L_0$)
is not suited for investigating the behavior of $\Delta P_z$ in the
resonant regime and in the epoch of oscillatory $\Delta P_z$
thereafter. Finally, we note that
the first calculation of the neutrino
asymmetry done by Enqvist {\sl et al.}\cite{Enq2}
shows an oscillatory asymmetry down to $T_6\sim 1$, because their
parameter choice, $\Delta m^2=-10^{-5}$ eV$^2$ and $\sin^22\theta=10^{-2}$,
does not satisfy eq.~(\ref{stillosc}).

\section{Implications}
We concentrate on two implications of a neutrino asymmetry as large as
eq.~(\ref{DeltaPzest}) in BBN.

The first is on other active-sterile neutrino oscillations in BBN
\cite{FTV,Foot}. When neglecting neutrino asymmetries, large areas of 
parameter space of active-sterile neutrino oscillation are ruled out
based on the argument that the sterile neutrino cannot be significantly
populated so as to violate the primordial $^4$He abundance observation.
The forbidden areas include the large angle $\numu$-$\nus$ mixing
with $\delta m^2\sim 10^{-2}$ eV$^2$ which can solve the atmospheric
neutrino problem, and the large angle $\nue$-$\nus$ mixing
with $\delta m^2\sim 10^{-5}$ eV$^2$ which can solve the solar neutrino
problem. This argument, however, no longer stands when a neutrino asymmetry
as large as in eq.~(\ref{DeltaPzest}) is in place.
For example, if $\nutau$ mixes with a lighter $\nus$ with
$\vert\delta m^2\vert\gg 10^{-2}{\rm eV}^2$ and an angle in
the shaded region of figure 3 (b),
the $\nutau\nutaubar$ asymmetry amplified by the $\nutau$-$\nus$ oscillation
can be large enough to suppress the $\numu$-$\nus$
oscillation from the $\numu$-$\nus$ mixing solution to the atmospheric problem
\cite{Foot}. Similarly, if $\nutau$ or $\numu$ mixes with a lighter $\nus$ with
$\vert\delta m^2\vert\gg 10^{-5}{\rm eV}^2$ and an angle in the shaded region
of figure 3 (b), the amplified neutrino asymmetry can be large enough to
suppress the $\nue$-$\nus$ oscillation originating from the
large angle $\nue$-$\nus$ mixing
solution to the solar neutrino problem. The suppression is in place even
in the epoch of oscillating $\Delta P_z$, because although an oscillating
$\Delta P_z$ constantly drives the other active-sterile neutrino
oscillations through resonance, the time of resonance crossing is too short
to allow any significant oscillation. Thus both these two solutions to the
atmospheric neutrino problem and the solar neutrino problem ruled
out previously may still be viable if an active neutrino
(more massive than $\numu$ or $\nue$ respectively) mixes with a
lighter sterile neutrino with parameters in the shaded region
of figure 3 (b).

The second implication is on the primordial $^4$He abundance itself.
Besides the number of neutrino species, the primordial $^4$He abundance
is also affected by a non-zero chemical potential
in the $\nue\nuebar$ sector at $T\sim 1$ MeV.
The mechanism is that the asymmetry in $\nue\nuebar$ changes
the neutron/proton conversion rates, thereby changes
the freeze-out time of the neutron to proton ratio as well as the
ratio itself. A $\xi_\nue$ (the $\nue$ chemical potential divided
by $kT$) of order 0.1 can induce an appreciable change in the prediction
of the primordial $^4$He abundance \cite{Reeves}.
When $\xi_\nue\ll 1$, $Y\approx Y(\xi_\nue=0)-0.234\xi_\nue$ \cite{Craig} and
$\Delta P_z=2(n_\nue-n_\nuebar )\approx 3.6\xi_\nue$. So
\begin{equation}
Y=Y_0-0.065\Delta P_z.
\end{equation}
The comparison with equation $Y=Y(N_\nu=3)+0.012(N_\nu-3)$
(where $N_\nu$ is the effective
number of neutrino species in BBN) \cite{Walker} indicates that
$\Delta P_z\sim 0.1$ in the $\nue\nuebar$ sector corresponds to roughly
$-0.55$ neutrino species, and therefore has a significant impact on the
predicted $^4$He abundance.

There are two ways to generate a $\xi_\nue$ of order $\pm$0.1
by active-sterile neutrino oscillations. The direct way is to have
a $\sim 1$ eV $\nue$ mix with a lighter $\nus$ (figure 3 (a)).
If the atmospheric neutrino problem and the solar neutrino problem are
to be solved by active neutrino oscillations, this implies that
all three active neutrinos are almost degenerate with a mass of order 1 eV.
This will be consistent with supernovae nucleosynthesis constraints
\cite{Qian,Sigl2} and compatible with the controversial
LSND result\cite{LSND1,LSND2} if the claimed detection of
$\numu$-$\nue$ oscillation solves the atmospheric neutrino problem
\cite{Fuller}.
Laboratory experiments limit the mass of $\nue$ to less than 5 eV
\cite{nuemass}. If $\nue$ is a majorana neutrino, its mass is further limited
to less than about 1 eV \cite{doublebeta}. 

The indirect way of generating a significant $\xi_\nue$
is to have $\nutau$ (or $\numu$) mix with a lighter $\nus$
with $10^2{\rm eV}^2\ga\vert\delta m^2\vert \ga 1$ eV$^2$
and a desired angle, and transfer
the asymmetry in the $\nutau\nutaubar$ ($\numu\numubar$) sector into
$\nue\nuebar$ by a $\nue$-$\nutau$ ($\numu$)
mixing. But to yield an asymmetry of order 0.1
as well in $\nue\nuebar$, the transfer has to be efficient. Take
the $\nue$-$\nutau$ oscillation as an example, the mixing has to satisfy
\begin{equation}
D^\prime\left({V_x^\prime\over V_z^\prime}\right)^2\ga H,
\end{equation}
where $D^\prime$, $V_x^\prime$ and $V_z^\prime$ denote the counterparts of
$D$, $V_x$ and $V_z$ in the $\nutau$-$\nus$ oscillation. Approximately
$$D^\prime\sim 5G_F^2T^5\approx T_6^3H,$$
$$V_x^\prime\approx {\delta M^2\over 6.3T}\sin 2\theta^\prime,$$
\begin{equation}
V_z^\prime\approx -{\delta M^2\over 6.3T}\cos 2\theta^\prime-180G_F^2T^5+
0.13G_FT^3(n_\nue-n_\nuebar)-0.13G_FT^3(n_\nutau-n_\nutaubar),
\end{equation}
where $\delta M^2$ and $\theta^\prime$ are the vacuum mixing parameters of
the $\nue$-$\nutau$ mixing. An efficient transfer of asymmetry means that
$(n_\nue-n_\nuebar )\sim (n_\nutau-n_\nutaubar )
\sim\vert\delta m^2\vert/18T_6^4$.
If the mass of $\nue$ is much lighter than $\nutau$, $\vert\delta M^2\vert
\approx\vert\delta m^2\vert$. Then $(D/H)(V_x^\prime/V_z^\prime)^2\sim 
T_6^3\sin^2 2\theta^\prime$ at temperatures approaching 1 MeV.
So if $\nutau$ has
a cosmologically interesting mass, $\Delta m^2 \sim 20$--100 eV$^{2}$,
$\vert\Delta P_z\vert\approx 0.1$ in $\nutau\nutaubar$
can be reached at $T\approx 2$--3 MeV by
the $\nutau$-$\nus$ mixing according to eq.~(\ref{DeltaPzest}),
which can efficiently transfer into an asymmetry of similar order in
$\nue\nuebar$ if $\nutau$ mixes with $\nue$ with $\sin^22\theta^\prime\sim 0.1$
and $\vert\delta M^2\vert\sim\vert\delta m^2\vert$.
This required mixing between $\nue$ and $\nutau$
lies near the edge of current lab limits on the $\nuebar$-$\bar\nu_x$
mixing \cite{Bugey}, and may be testable in the near future.
Based on supernovae nucleosynthesis arguments, however,
the required $\nue$-$\nutau$ mixing is ruled out
(although supernova models are uncertain to some extent) \cite{Qian,Sigl2}.
The above analyses applies similarly to the $\numu$-$\nus$ and
$\numu$-$\nue$ mixings, but the required mixing between $\nue$ and $\numu$
to transfer asymmetries efficiently has already
been ruled out by laboratory experiments \cite{nuemass}.

Of course, the indirect way of transferring asymmetries from $\nutau\nutaubar$
(or $\numu\numubar$) into $\nue\nuebar$ works if they have almost degenerate
masses, $\sim 1$eV. In this case the required $\nue$-$\numu$ ($\nutau$)
mixing will not be ruled out by
laboratory experiments or astrophysical considerations.

Recently, Hata et al. \cite{Hata} suggested a possible ``crisis''
in BBN because the predicted $^4$He abundance from the standard BBN, coupled
with predictions of $^3$He and D from generic chemical evolution
models, is too high to be consistent
with the observed abundance, unless the number of effective neutrino species 
$N_\nu$ at the time of BBN is less than 2.6.
This is at odds with popular beliefs and
most theoretical models which assume all $\nue$, $\numu$ and $\nutau$ to be
lighter than $\sim 1$ MeV and therefore $N_\nu\ge 3$.
But as seen above, if an asymmetry of order $\Delta P_z\sim
0.1$ arises in the $\nue\nuebar$ sector from active-sterile neutrino
oscillations, this potential ``crisis'' can be solved. Of course,
the mixing parameters of the active-sterile oscillations have to be right
to yield a positive $\Delta P_z$ instead of a negative one. An interesting
note is that the neutrino mass required to solve this ``crisis''
is $\sim 1$ eV, which qualifies neutrinos as dark matter candidates.
A $\nue$ mass of $\sim 1$ eV is
within a factor of 5 below the current limit on $\nue$ mass, and
right on the edge of detection limit if $\nue$ are majorana neutrinos.
If not introducing more sterile neutrinos, the solar neutrino problem
and the atmospheric neutrinos have to be solved by mixings among the three
active neutrinos, therefore requiring their masses to be almost degenerate.
A model of three neutrinos almost degenerate in mass is favorable in forming
structures in the universe \cite{Primack}, but may not be the most natural
theoretical model so far.

\section{Summary}

In summary, we have calculated the parameter space of active-sterile neutrino
mixings that amplifies neutrino chemical potentials,
and the size of the amplification. Results
are summarized in figure 3. By exploring the sensitivity of the
amplification to the initial condition, the mixing parameters and the
calculational parameters, and by analyzing the Lyapunov exponent of the
system, we showed evidences that the amplification process
is chaotic. We have also discussed the implications of our results
on BBN. It was shown that a $\nue$ chemical potential
of order 0.1$kT$ could be achieved by either
a mixing between $\sim 1$ eV $\nue$ and a lighter sterile neutrino,
or a mixing between a $\sim 1$ eV $\numu$ (or $\nutau$)
and a lighter sterile neutrino coupled with a mixing between
almost degenerate $\nue$ and $\numu$ (or $\nutau$). Such a chemical
potential in $\nue$ can lower the effective
number of neutrino species to significantly below 3.

\section{Acknowledgment}
The author thanks Martin Duncan, Man-Hoi Lee and G\"unter Sigl for helpful
discussions. The author also thanks Craig Copi and David Schramm
for providing a primordial nucleosynthesis code.
This work is supported by NSERC grant of Canada and CITA National Fellowship.

\newpage
\noindent{\bf Figure Captions:}
\bigskip

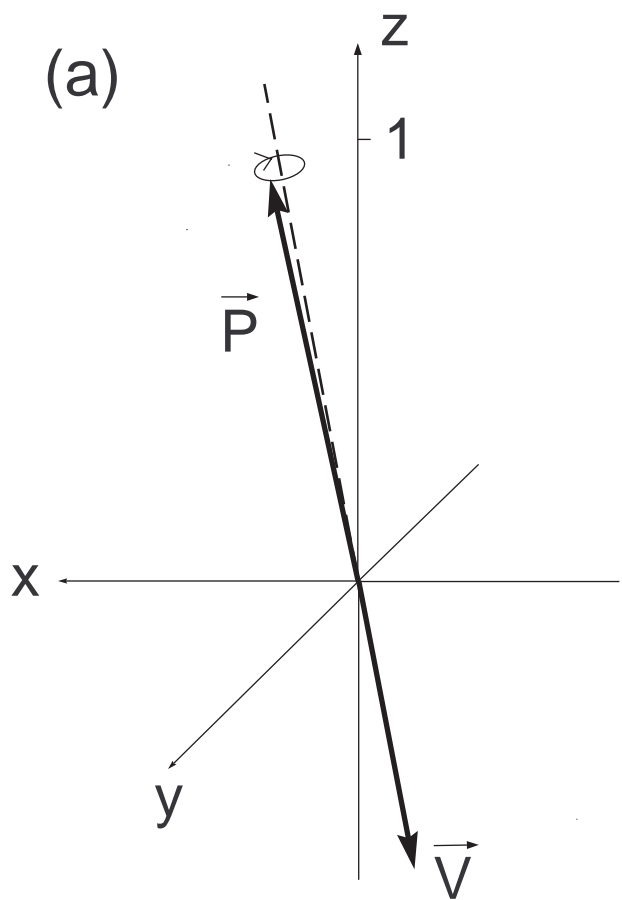
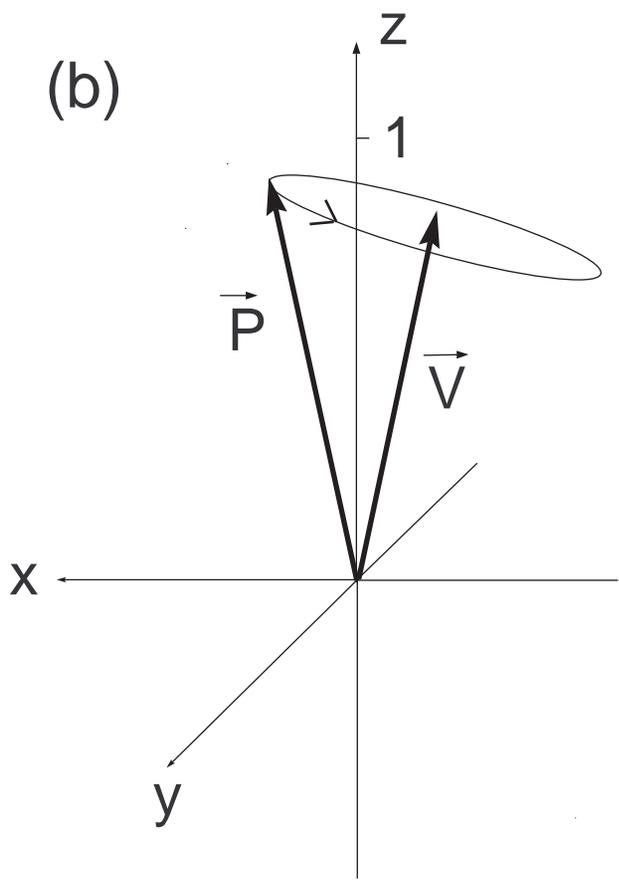
\noindent Figure 1. Illustrations of evolution of {\bf P} and {\bf V}
for $\delta m^2<0$. (a) before resonance; (b) right after resonance.
\bigskip

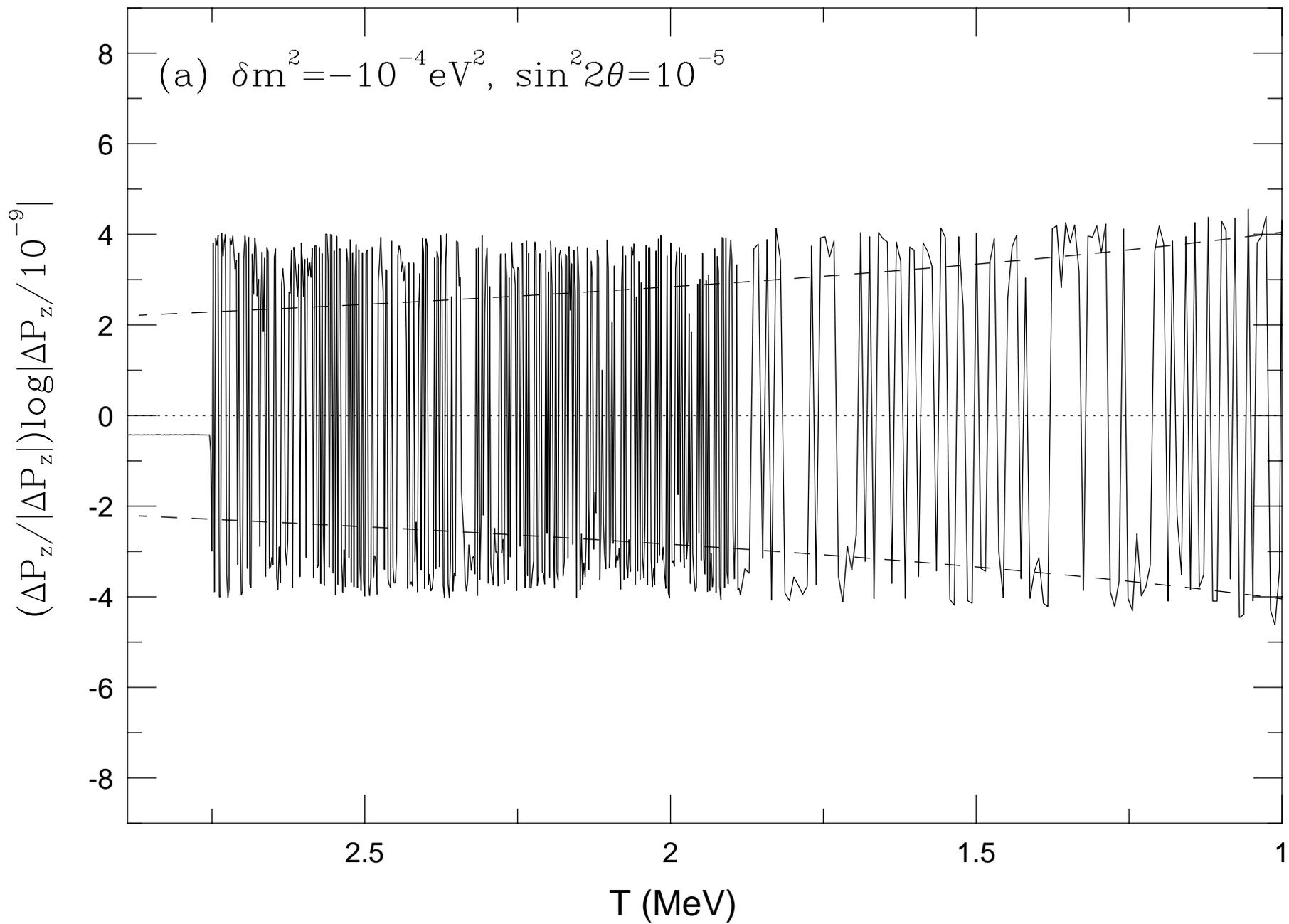
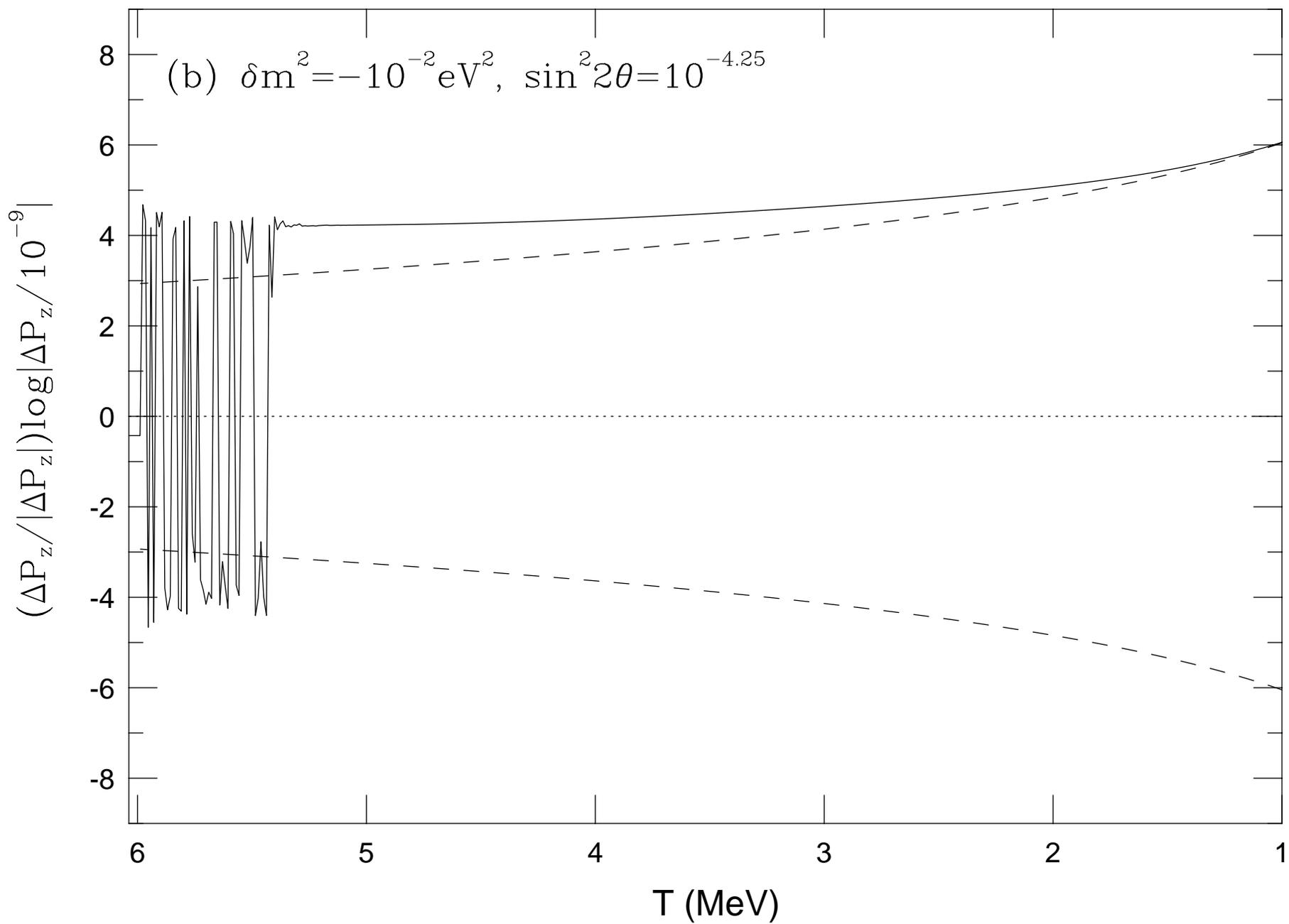
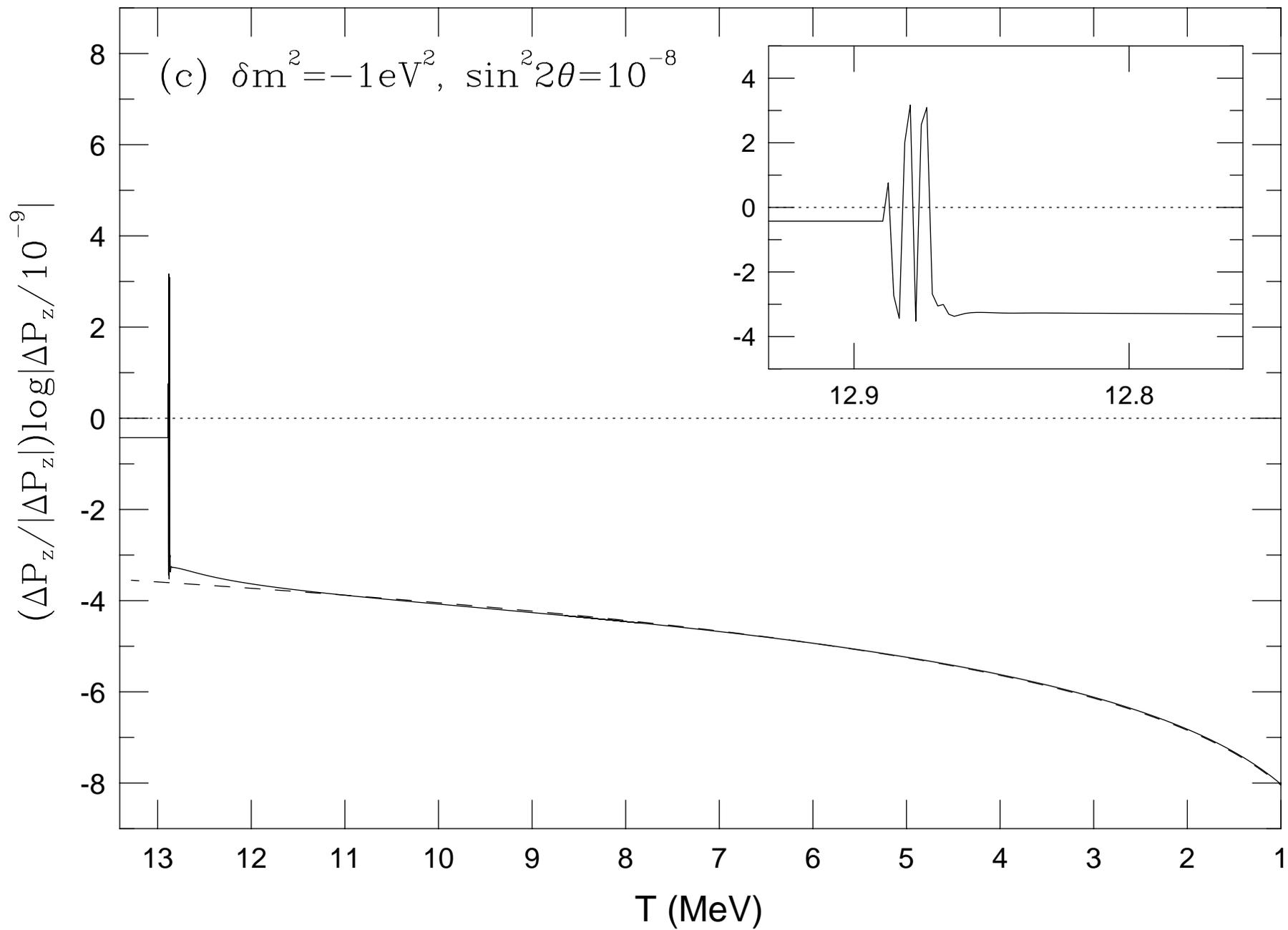
\noindent Figure 2. The solid lines show the evolution of $\Delta P_z$ vs.
the temperature of the universe, for $\nue$-$\nus$ mixing. Asymmetries
less than $10^{-9}$ are ignored. The
dash line shows $\vert\Delta P_z\vert=\vert\delta m^2/9{\rm eV}^2\vert
T_6^{-4}$. $L_0=10^{-9}$.
(a) $\delta m^2=-10^{-4}$ eV$^2$, $\sin^22\theta=10^{-5}$;
(b) $\delta m^2=-10^{-2}$ eV$^2$, $\sin^22\theta=10^{-4.25}$;
(c) $\delta m^2=-1$ eV$^2$, $\sin^22\theta=10^{-8}$.
\bigskip

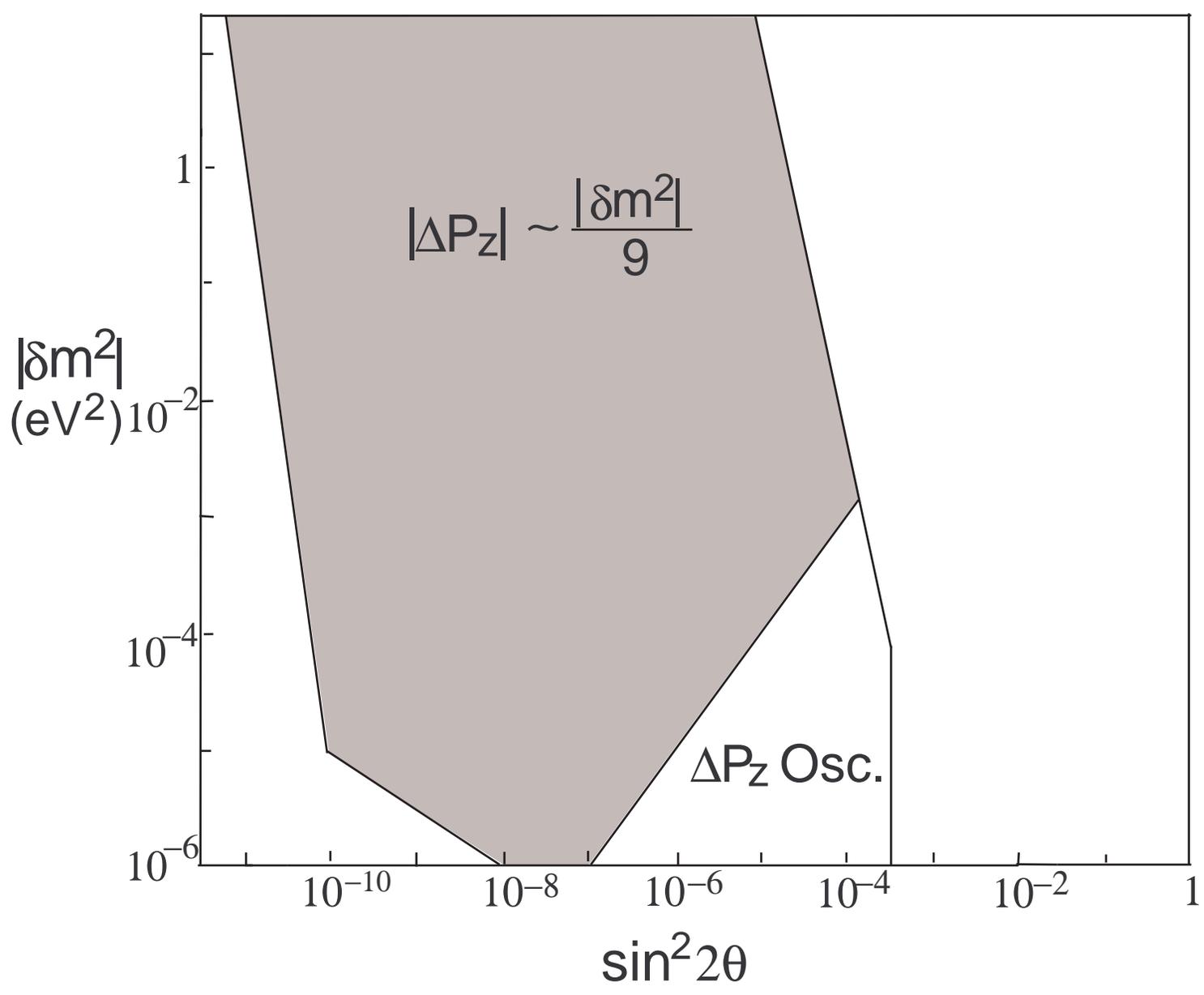
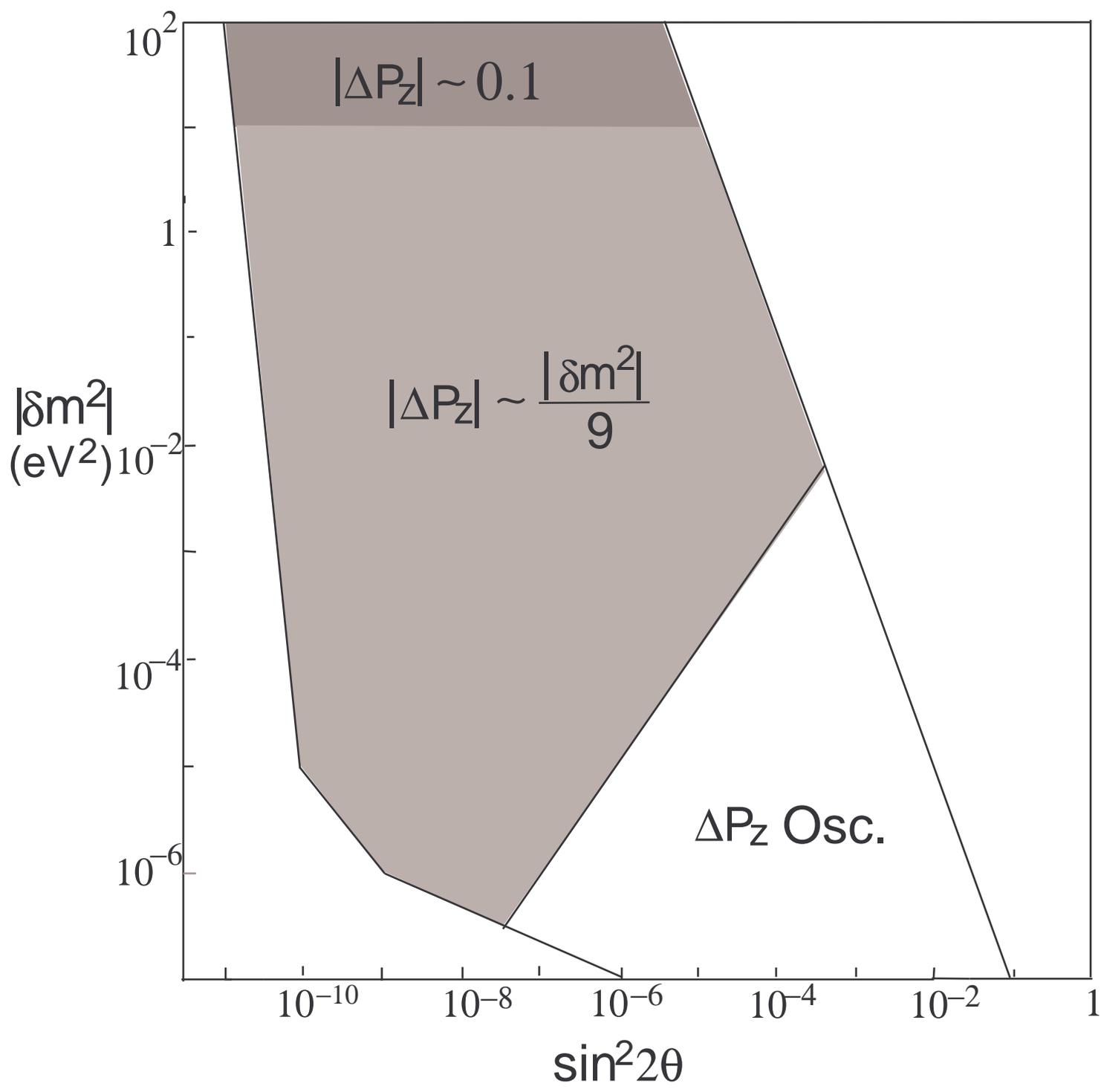
\noindent Figure 3. Regions in between the two thick lines represent
allowed mixing parameters that amplify neutrino chemical potential. The
magnitude of $\Delta P_z$ at 1 MeV is shown. The region at the lower
right noted with ``$\Delta P_z$ Osc.'' has an oscillating $\Delta P_z$
at 1 MeV. For $\vert\delta m^2\vert\ga 10$ eV$^2$, $\Delta P_z$ is
limited to $\sim 0.1$.
(a) $\nue$-$\nus$ mixing ($\nue$ heavier than $\nus$); 
(b) $\numu$($\nutau$)-$\nus$ mixing ($\numu$ or $\nutau$ heavier than $\nus$).

\end{document}